\def\be{\begin{equation}}
\def\ee{\end{equation}}
\def\bea{\begin{eqnarray}}
\def\eea{\end{eqnarray}}
\def\bse{\begin{subequations}}
\def\ese{\end{subequations}}

\def\Hcone{H_{\text{c}1}}
\def\Hctwo{H_{\text{c}2}}
\documentclass[prl,twocolumn,showpacs,amsmath,amssymb,preprintnumbers]{revtex4}
\usepackage{graphicx}
\usepackage{dcolumn}
\usepackage{bm}
\begin{document}
\preprint{Phys. Rev. Lett. {\bf 98}, 187002 (2007)}
\title{Elasticity and melting of skyrmion flux lattices in $p\,$-wave superconductors
}
\author{Qi Li$^{1,2}$, John Toner$^{1}$, and D. Belitz$^{1,2}$}
\affiliation{$^{1}$Department of Physics and Institute of Theoretical Science,
              University of Oregon, Eugene, OR 97403\\
          $^{2}$Materials Science Institute, University of Oregon, Eugene,
OR 97403
          }
\date{\today}
\begin{abstract}
We analytically calculate the energy, magnetization curves ($B(H)$), and
elasticity of skyrmion flux lattices in $p\,$-wave superconductors near the
lower critical field $\Hcone$, and use these results with the Lindemann
criterion to predict their melting curve. In striking contrast to vortex flux
lattices, which {\it always} melt at an external field $H > \Hcone$, skyrmion
flux lattices {\it never} melt near $\Hcone$. This provides a simple and
unambiguous test for the presence of skyrmions.
%
\end{abstract}


\maketitle

The topological excitations known as skyrmions have been proposed to have many
applications. After they were introduced in a nuclear physics context by Skyrme
\cite{Skyrme_1961}, variations of this concept have been shown or proposed to
be important in superfluid $^3$He \cite{Anderson_Toulouse_1977}, in quantum
Hall systems \cite{Timm_Girvin_Fertig_1998}, in $p\,$-wave superconductors
\cite{Knigavko_Rosenstein_Chen_1999}, and in metallic magnets
\cite{Roessler_Bogdanov_Pfleiderer_2006}. Given this widespread predicted
occurrence, it is desirable to find a simple experimental signature that can
serve as a smoking gun indicating the presence of skyrmions. In this Letter we
show that, in $p\,$-wave superconductors subject to an external magnetic field,
the structure of the phase diagram provides an unambiguous test for the
presence of skyrmions \cite{p-wave_footnote}.

In superconductors, skyrmions compete with another, and more well-known,
species of topological excitations, viz., vortices. In type-II superconductors,
the latter are induced by an external magnetic field $H$ and form the famous
Abrikosov flux lattice in a field range $\Hcone < H < \Hctwo$. It is known both
theoretically\cite{Huberman_Doniach_1979, Fisher_1980, Nelson_Seung_1989,
Brandt_1989} and experimentally \cite{Gammel_et_al_1988, Safar_et_al_1992} that
these flux lattices can melt. They do so near both $\Hcone$ and $\Hctwo$, where
the elastic constants of the flux lattice vanish, which in clean systems makes
the root-mean-square positional thermal fluctuations $\sqrt{\langle\vert{\bm
u}({\bm x})\vert^2\rangle}$ grow without bound as these fields are approached.

Vortices involve a singular field configuration at their cores and can occur
for any symmetry of the superconducting order parameter. In $p\,$-wave
superconductors the vector character of the superconducting order parameter
also allows for skyrmions, which, in contrast to vortices, are {\em
non-singular} topological defects. Like vortices, each skyrmion carries a
quantized magnetic flux, and for strongly type-II $p\,$-wave superconductors in
a field $\Hcone < H < \Hctwo$, a skyrmion lattice is predicted to have a lower
energy than a vortex lattice \cite{Knigavko_Rosenstein_Chen_1999}. The
currently most convincing case for a $p\,$-wave superconductor is Sr$_2$RuO$_4$
\cite{Nelson_et_al_2004}, another candidate is UGe$_2$
\cite{Machida_Ohmi_2001}.

\begin{figure}[t,h]
\vskip -1mm
\includegraphics[width=7cm]{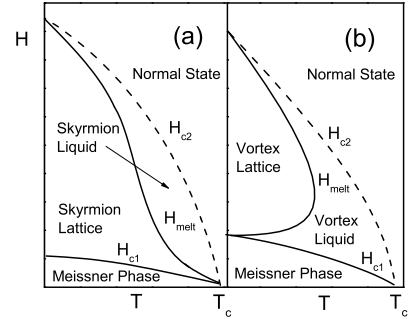}
\vskip -0mm
\caption{(a) External field ($H$) vs. temperature ($T$) phase diagram for
skyrmion flux lattices. There is a direct transition from the skyrmion flux
lattice to the Meissner phase. The theory predicts the shape of the melting
curve only close to $T_{\text{c}}$; the rest of the curve is an educated guess.
(b) Same as (a) for vortex flux lattices. The vortex flux lattice always melts
before $H$ reaches $\Hcone$.}
\label{fig:1}
\end{figure}

Our central result is the prediction that one can distinguish a skyrmion
lattice from a vortex lattice by considering the melting curve of the lattice.
Our results for the latter are summarized in the phase diagram for clean
skyrmion flux lattices shown in Fig.\ \ref{fig:1}(a). This is strikingly
different from that for the vortex case, Fig.\ \ref{fig:1}(b): The skyrmion
lattice {\em never} melts near $\Hcone$, while vortex lattices {\em always}
melt near $\Hcone$.

The physics behind this result is simple. The interaction between vortices
falls off exponentially at large distances \cite{Tinkham_1975}. As a result,
when $H \to \Hcone$ from above, which causes the vortex lattice constant $R$ to
grow, the elastic constants vanish exponentially with $R$ as $R\to\infty$,
which, in turn, causes the root-mean-square displacement fluctuation
$\sqrt{\vert{\bm u}({\bm x})\vert^2}$ to diverge exponentially.
Consequently, $\sqrt{\vert{\bm u}({\bm x})\vert^2} \gg R$ as $R\to\infty$
$H\to\Hcone$. The Lindemann criterion \cite{Chaikin_Lubensky_1995} then implies
that the vortex lattice {\it must} melt with decreasing $H$ before one reaches
$\Hcone$.

For skyrmions, as first shown numerically in Ref.\
\onlinecite{Knigavko_Rosenstein_Chen_1999}, and confirmed analytically below,
the interaction potential falls off only as $1/R$. As a result, the Lam{\'e}
coefficients $\mu$ and $\lambda_{\text{L}}$ of the skyrmion lattice vanish only
as $1/R^3$ as $R\to\infty$. This leads to $\sqrt{\vert{\bm u}({\bm x})\vert^2}
\propto R^{3/4} \ll R$. Hence, the skyrmion lattice {\it never} melts as
$\Hcone$ is approached. The shape of the phase diagram alone thus distinguishes
skyrmion lattices from vortex lattices.

All of these results follow from an analytic solution of the Euler-Lagrange
equations that give the minimum energy configuration for a skyrmion lattice in
the $\beta$-phase of a $p\,$-wave superconductor. We have obtained an
asymptotic solution in the limit of large lattice spacing $R \gg \lambda$,
where $\lambda$ is the London penetration depth. Near $\Hcone$ the
superconductor is always in this limit. As in the numerical work of Ref.\
\onlinecite{Knigavko_Rosenstein_Chen_1999}, we replace the hexagonal unit cell
one expects in the skyrmion lattice by a circle of the same area. We expect
this approximation to preserve the correct scaling of the energy.

\begin{figure}[t]
\vskip -0mm
\includegraphics[width=8cm]{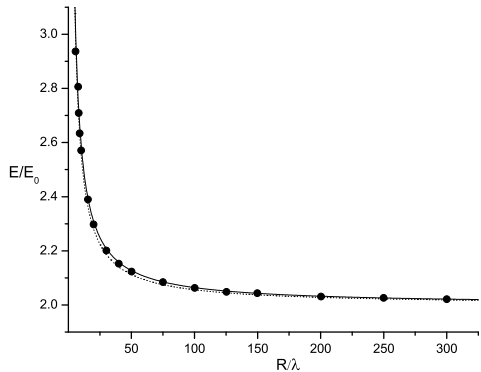}
\vskip -0mm
\caption{Numerical data for the energy per skyrmion per unit length (circles)
together with the best fit to a pure $1/R$ behavior (dashed line) from Ref.\
\cite{Knigavko_Rosenstein_Chen_1999}, and the analytic solution given by Eq.
(\ref{eq:1}) (solid line).}
\label{fig:2}
\end{figure}

For the first three terms of the asymptotic expansion of the energy $E$ per
skyrmion per unit length we obtain
\be
E/E_0 = 2 + (8\sqrt{6}/3)\,(\lambda/R) -
(16/3)\,(\lambda/R)^2\,\ln(R/\lambda).\!\!
\label{eq:1}
\ee
Here $E_0 = (\Phi_0/4\pi\lambda)^2$, with $\Phi_0 = \pi\hbar c/e$ the flux
quantum. This analytic result is in excellent agreement with the numerical
solution reported in Ref. \cite{Knigavko_Rosenstein_Chen_1999}, see Fig.\
\ref{fig:2}. Notice that there are no free parameters in Eq. (\ref{eq:1}).

In the remainder of this Letter we sketch the derivation of the above results.
A complete account of the calculations will be given elsewhere \cite{us_tbp}.
We begin by reviewing the formulation of the skyrmion lattice problem
\cite{Knigavko_Rosenstein_Chen_1999}.

The spin part of the order parameter for a $p\,$-wave superconductor is a
complex 3-vector ${\bm \psi}({\bm x})$. In a large regime of Landau theory
parameter space (the so-called $\beta$-phase), all low-energy configurations of
${\bm \psi}({\bm x})$ can be written in the form ${\bm \psi}({\bm x}) = \psi_0
({\hat {\bm n}}({\bm x})+i{\hat {\bm m}}({\bm x}))$, where ${\hat{\bm n}}({\bm
x})$ and ${\hat{\bm m}}({\bm x})$ are real, mutually orthogonal unit vectors
that vary slowly in space, and $\psi_0$ is a constant
\cite{Knigavko_Rosenstein_Chen_1999}.

In the ground state, ${\hat {\bm n}}$ and ${\hat {\bm m}}$ are constants. Slow
spatial variations ${\hat {\bm n}}({\bm x})$ and ${\hat {\bm m}}({\bm x})$ cost
an energy \cite{crystal_field_footnote}
\bea
H_{\text{L}} &=& \int d{\bm x}\ \Bigl[\frac{1}{2}\,(\partial_i {\hat {\bm
l}})^2 + ({\hat {\bm n}}\cdot\partial_i{\hat {\bm m}} - {\bm a})^2 +
({\bm\nabla}\times{\bm a})^2 \nonumber\\
&&\hskip 30pt - 2{\bm h} \cdot ({\bm\nabla} \times {\bm a}) \Bigr].
\label{eq:2}
\eea
Here we use dimensionless units where distance, vector potential ${\bm a}$,
magnetic field ${\bm h}$, and energy are measured in units of $\lambda$,
$\Phi_0/2\pi\lambda$, $\Phi_0/2\pi\lambda^2$, and $\Phi_0^2/32\pi^3\lambda$,
respectively \cite{Knigavko_Rosenstein_Chen_1999}. ${\hat {\bm l}} \equiv
{\hat{\bm n}}\times {\hat{\bm m}}$, and the last term in the London energy, Eq.
(\ref{eq:2}), represents the coupling of the external magnetic field ${\bm h}$
to the magnetic induction ${\bm b} = {\bm\nabla}\times{\bm a}$.

A vortex is a low-energy configuration of ${\hat{\bm n}}({\bm x})$ and
${\hat{\bm m}}({\bm x})$ in which ${\hat{\bm l}}$ is a constant (i.e., in which
${\hat{\bm n}}({\bm x})$ and ${\hat{\bm m}}({\bm x})$ span the same plane for
all ${\bm x}$), but ${\hat {\bm n}}({\bm x})$ and ${\hat{\bm m}}({\bm x})$
rotate by $2\pi n$ ($n$ integer) as one follows their evolution around any
closed spatial path that encircles the path of the vortex core. Such a
configuration necessarily has a singularity on the path of the core.

The coupling between ${\hat {\bm n}}({\bm x})$, ${\hat {\bm m}}({\bm x})$, and
${\bm a}$ makes the formation of vortices energetically favorable when ${\bm h}
= (0,0,h)$ is sufficiently strong. In conventional type-II superconductors,
this leads to the appearance of a spontaneous vortex flux lattice, where ${\bm
b}$ is parallel to ${\bm h}$, and ${\hat {\bm n}}({\bm x})$, ${\hat {\bm
m}}({\bm x})$, and ${\bm b}({\bm x})$ all become spatially periodic functions
of $x$ and $y$ (${\bm x} = (x,y,z))$. This two-dimensional spatially periodic
structure forms a hexagonal lattice.

For $s$-wave superconductors, vortices are the only possible topological
defects. For $p\,$-wave superconductors, the 3-d freedom of the vectors ${\hat
{\bm n}}({\bm x})$ and ${\hat {\bm m}}({\bm x})$ also allows for skyrmions,
which are {\it non-singular} configurations of ${\hat {\bm n}}({\bm x})$ and
${\hat {\bm m}}({\bm x})$. They look like $n=2$ vortices at the unit cell
boundary. Inside the cell, ${\hat {\bm n}}({\bm x})$ and ${\hat {\bm m}}({\bm
x})$ move out of their common plane, so ${\hat {\bm l}}(\bm x)$ is no longer
constant. In the simplest case the skyrmion is cylindrically symmetric, and
${\hat {\bm l}}(\bm x)$ forms an angle $\theta(\bm x)$ with some central axis
(which we will take to be the $z$-axis). While $\theta(\bm x)$ changes from $0$
to $\pi$ as one moves from infinity to the skyrmion axis, ${\hat {\bm n}}({\bm
x})$ and ${\hat {\bm m}}({\bm x})$ rotate around $\hat{\bm l}$ by $4\pi$ on any
loop enclosing the skyrmion axis.

This picture leads to the characterization of a cylindrically symmetric
skyrmion in terms of the single, as yet undetermined, angle $\theta(\bm x)$. In
polar coordinates $(r,\varphi)$ the vector fields ${\hat {\bm n}}({\bm x})$,
${\hat {\bm m}}({\bm x})$ and ${\hat {\bm l}}(\bm x)$ are given by
\bea
{\hat {\bm l}} &=& {\hat {\bm e}}_z \cos\theta(r) + {\hat{\bm e}}_r\sin
\theta(r),
\nonumber\\
{\hat{\bm n}} &=& \left({\hat{\bm e}}_z\sin\theta(r) - {\hat{\bm e}}_r\cos
\theta (r)\right) \sin\varphi + {\hat{\bm e}}_{\varphi}\cos \varphi,
\nonumber\\
{\hat{\bm m}} &=& \left({\hat{\bm e}}_z \sin\theta(r) - {\hat{\bm e}}_r\cos
\theta(r)\right) \,\cos \varphi - {\hat{\bm e}}_{\varphi}\sin \varphi.
\label{eq:3}
\eea
If the field $\theta(\bm x)$ minimizes the energy of this configuration, then
that configuration is a local minimum of the London energy, Eq. (\ref{eq:2})
\cite{Knigavko_Rosenstein_Chen_1999}.

Inserting Eq. (\ref{eq:3}) into Eq. (\ref{eq:2}) yields the energy per unit
length, in units of $E_0$, of a cylindrically symmetric skyrmion in a region of
radius $R$,
\bea
E &=& \frac{1}{2} \int_0^R dr\,r\left[\left(\theta'(r)\right)^2
       + \frac{1}{r^2}\,\sin^2 \theta(r)\right]
\nonumber\\
&&+ \int_0^R dr\,r\left[\frac{1}{r}\,\left(1 + \cos\theta(r)\right) +
a(r)\right]^2
\nonumber\\
&&+ \int_0^R dr\,r \left[a(r)/r + a'(r)\right]^2.
\label{eq:4}
\eea
The three terms represent the energy of a nonmagnetic skyrmion, the
supercurrent energy, and the magnetic energy, respectively. Skyrmions in a
lattice are {\it not} cylindrically symmetric, since the lattice is not.
However, since a hexagon is well approximated by a circle of the same area, we
follow Ref.\ \onlinecite{Knigavko_Rosenstein_Chen_1999} by so approximating the
unit cell.

The Euler-Lagrange equations that result from minimizing the above energy read
\bse
\label{eqs:5}
\bea
\theta''(r) + \frac{1}{r}\,\theta'(r) =
\frac{-\sin\theta(r)}{r}\,\left[\frac{2+\cos\theta(r)}{r} + 2 a(r)\right],\
\label{eq:5a}\\
a''(r) + \frac{1}{r}\,a'(r) - \frac{1}{r^2}\,a(r) = a(r) + \frac{1}{r}\,\left[1
+ \cos\theta(r)\right].\quad
\label{eq:5b}
\eea
\ese
This set of coupled, nonlinear ODEs must be solved subject to the boundary
condition $\theta(r=0) = \pi$, and $\theta(r=R) = 0$. In Ref.\
\onlinecite{Knigavko_Rosenstein_Chen_1999} this was done numerically using
finite elements methods. Here we show that for large $R$ (i.e., near the lower
critical field $\Hcone$) an analytic solution can be obtained in the form of an
asymptotic expansion in powers of $1/R$. To zeroth order, i.e, for $R \to
\infty$, the l.h.s. of Eq. (\ref{eq:5b}) vanishes, so $a$ is given by
\bse
\label{eqs:6}
\be
a_{\infty}(r) = -\,\left[1 + \cos\theta(r)\right]/r.
\label{eq:6a}
\ee
Inserting this in Eq. (\ref{eq:5a}), the resulting ODE is solved by
\be
\theta_{\infty}(r) = 2\arctan(\ell/r),
\label{eq:6b}
\ee
\ese
for any $\ell$. All of these solutions fulfill the boundary condition, and
$\ell$ is undetermined at this point.

We write $\theta(r) = \theta_{\infty}(r) + \delta\theta(r)$ and $a(r) =
a_{\infty}(r) + \delta a(r)$, and require $\vert\delta\theta(r)\vert \ll 1$ and
$\vert \delta a(r)\vert \ll \vert 1 + \cos\theta_{\infty}(r)\vert/r$. We do
{\em not} require $\vert\delta\theta\vert \ll \theta_{\infty}$, which is
crucial for the success of our perturbative method. Inserting this into the
ODEs, one sees that $\delta\theta$ can be written
\be
\delta\theta(r) = (1/\ell^2)g(r/\ell) + (1/\ell^4)h(r/\ell) +  O(1/\ell^6),
\label{eq:7}
\ee
while $\delta a$ can be expressed in terms of $\theta_{\infty}$,
$\delta\theta$, and its derivatives. It will turn out that $\ell \propto
\sqrt{R}$, so this is the desired expansion in powers of $1/R$. For the
function $g$ we obtain a linear, inhomogeneous ODE of second order,
\be
g''(u) + \frac{1}{u}\,g'(u) - \frac{u^4 - 6u^2 + 1}{u^2(1+u^2)^2}\,g(u) =
\frac{-64u}{(1+u^2)^4},
\label{eq:8}
\ee
The physical solution is the one that diverges linearly for large arguments.
Standard methods 
give
\be
g(u) = \frac{-4}{3}\,\frac{u[u^2(4+u^2) + 2(1+u^2)\ln(1+u^2)]}{(1+u^2)^2}\ .
\label{eq:9}
\ee
The parameter $\ell$ is now determined by the requirement $\theta(r=R)=0$,
which yields, to this order, $\ell = (2/3)^{1/4}\,R^{1/2}$. To the same order,
$\delta a$ is just given by the l.h.s. of Eq. (\ref{eq:5b}) with $a$ replaced
by $a_{\infty}$,
\be
\delta a(r) = (16 r/\ell^4)(1 + r^2/\ell^2)^{-3} + O(1/\ell^5).
\label{eq:10}
\ee
Inserting these results in Eq. (\ref{eq:4}) and performing the integrals yields
the first two terms on the r.h.s. of Eq. (\ref{eq:1}). This method can be
continued order by order. At the next order it yields the final term in Eq.
(\ref{eq:1}).

From the energy per skyrmion, Eq. (\ref{eq:1}), we can calculate the external
field dependence of the equilibrium lattice constant, $R(H)$. This is done by
minimizing the energy per unit volume, which is the energy per unit length per
skyrmion, Eq. (\ref{eq:1}), divided by the area per skyrmion, $\pi R^2$, minus
an energy density {\em gain} of $2 \Phi_0 H/\pi R^2$ due to the external field.
The latter is obtained from the ${\bm h} \cdot ({\bm \nabla} \times {\bm a})$
term in Eq. (\ref{eq:2}) by noting that the magnetic flux $\int dx\,dy\, {\hat
{\bm z}}\cdot({\bm \nabla} \times {\bm a}) = 2 \Phi_0$ for each skyrmion in the
lattice. This yields the Gibbs free energy density
\be
g(R) = (K/4\pi^2)\left[-\delta/R^2 + 4\sqrt{6}\lambda/(3R^3)\right],
\label{eq:11}
\ee
where $K \equiv \Phi_0^2/2\pi\lambda^2$ and $\delta\equiv H/\Hcone - 1$ with
$\Hcone \equiv K/2\Phi_0$.

For $H < \Hcone$, $g(R)$ is minimized by $R \to \infty$; i.e., there are no
skyrmions. For $H > \Hcone$, on the other hand, the energy density is minimized
at a finite R given by
\be
R = R_0 \equiv 2\sqrt{6}\lambda/\delta.
\label{eq:12}
\ee
Thus, $\Hcone$ is the lower critical field at which the skyrmion lattice first
forms. This implies for the spatially averaged magnetic induction, which is the
flux per skyrmion $2 \Phi_0$ divided by the unit cell area,
\be
B(H) = 2\Phi_0/(\pi R_0^2) = \delta^2 \Hcone/3.
\label{eq:13}
\end{equation}
Hence, $B(H)$ is {\em horizontal} near $\Hcone$, while it is {\em vertical} for
vortex lattices. This result, with a slightly different numerical prefactor,
was first obtained numerically in \cite{Knigavko_Rosenstein_Chen_1999}.

We now calculate the elastic properties of the skyrmion lattice. By symmetry,
the elastic Hamiltonian of a hexagonal lattice of lines parallel to the
$z$-axis in 3-d is
\be
H_{\text{el}} = \frac{1}{2}\int d{\bm x}\ \left(2\mu\, u_{\alpha\beta}\,
u_{\alpha\beta} + \lambda_{\text{L}}\, u_{\alpha\alpha}\, u_{\beta\beta} +
K_{\text{tilt}}\vert\partial_z{\bm u}\vert^2\right),
\label{eq:14}
\ee
where $u_{\alpha \beta}\equiv (\partial_{\beta} u_{\alpha} + \partial_{\alpha}
u_{\beta})/2$ is the strain tensor, $\alpha,\beta \in \{x,y\}$, and $\bm u$
only has $x$ and $y$ components. $\mu$, $\lambda_{\text{L}}$, and
$K_{\text{tilt}}$ are the shear, bulk, and tilt moduli.

Consider the energy change due to a dilation of the lattice, $R_0 \to
R_0^{\,\epsilon} \equiv R_0(1+\epsilon)$ with $\epsilon \ll 1$. This
corresponds to a displacement ${\bm u}(\bm x) = \epsilon\, {\bm x}_{\perp}$,
and a strain tensor $u_{\alpha
\beta } = \epsilon\, \delta^{\perp}_{\alpha\beta}$, with ${\bm x}_{\perp}$ the
projection of ${\bm x}$ perpendicular to ${\hat {\bm z}}$. 
Inserting this in $H_{\text{el}}$ gives $E_{\text{dil}}/V = 2(\mu +
\lambda_{\text{L}})\,\epsilon^2$, and comparing with the dilation energy
implied by Eq. (\ref{eq:11}),
\[
\frac{E_{\text{dil}}}{V} = g\left(R_0^{\,\epsilon}\right) - g(R_0) =
\frac{1}{2}\,g''(R_0)\,(\epsilon R_0)^2 = \frac{K \delta^3}{96 \pi^2
\lambda^2}\ \epsilon^2.
\]
gives
\be
\mu +\lambda_{\text{L}} = K \delta^3/192 \pi^2\lambda^2 .
\label{eq:15}
\ee

Obtaining $\mu$ is more difficult since we have approximated the unit cell by a
circle. We use a heuristic approach. Equation (\ref{eq:11}) is of the form that
would result if the skyrmions interacted via a nearest-neighbor pair potential
$U_p(r)\propto K \lambda/r$. Pretending that $U_p(r)$ is the origin of the
skyrmion energy (which should give the correct scaling of $\mu$ with $h$) makes
it straightforward to calculate $\mu$, and comparing the result with Eq.
(\ref{eq:15}) yields $\lambda_{\text{L}}$. We obtain
\be
\mu \sim \lambda_{\text{L}} = K \delta^3/\lambda^2 \times O(1).
\label{eq:16}
\ee
$K_{\text{tilt}}$ we obtain by considering a uniform tilt of the skyrmion axes
away from the $z$-axis by a small angle $\theta = \vert\partial_z{\bm u}\vert$.
The tilt energy in Eq. (\ref{eq:14}) is the change in the $\bm b \cdot \bm h$
energy in Eq. (\ref{eq:2}). Since, per skyrmion per unit length, this energy is
(in ordinary units) $-\Phi_0 H \cos\theta/2\pi$, tilting {\it changes} this
energy by $\Phi_0 H (1 - \cos\theta)/2\pi \approx \Phi_0 H \theta^2/4\pi =
\Phi_0 H \vert\partial_z{\bm u}\vert^2/4\pi$. Dividing this result by the unit
cell area, using Eq. (\ref{eq:12}) for $R_0$, and identifying the result with
the $K_{\text{tilt}}$ term in Eq. (\ref{eq:14}) gives $K_{\text{tilt}}$ near
$\Hcone$:
\be
K_{\text{tilt}} = (\Hcone \delta)^2/12\pi .
\label{eq:17}
\ee

We can now calculate the mean-square positional fluctuations ${\langle\vert{\bm
u}({\bm x})\vert^2\rangle}$ by Fourier transforming Eq. (\ref{eq:14}) and using
the equipartition theorem. This gives
\be
\langle\vert{\bm u}({\bm x})\vert^2\rangle_{\text{T}} = \frac{k_{\text{B}}T}{V}
\sum_{\bm q \in \text{BZ}} 1/(\mu\,q_{\perp}^2 + K_{\text{tilt}}\,q_z^2)
\label{eq:18}
\ee
for the transverse fluctuations, and the same expression with $\mu \to (2\mu +
\lambda_{\text{L}})$ for the longitudinal ones. $q_\perp$ and $q_z$ are the
projections of $\bm q$ orthogonal to and along $z$, respectively. The
Brillouin zone (BZ) of the skyrmion lattice is a hexagon of edge length
$\propto 1/R_0$ in the plane perpendicular to $z$, and infinitely extended in
the $z$-direction.

Since $\mu$ and $\lambda_{\text{L}}$ are comparable in magnitude, the same is
true for the longitudinal and the transverse contributions to $\langle\vert{\bm
u}({\bm x})\vert^2\rangle$. Performing the integral over $q_z$ yields
\be
\langle\vert{\bm u}({\bm x})\vert^2\rangle \sim \langle\vert{\bm u}({\bm
x})\vert^2\rangle_{\text T} = \int_{\text{BZ}} \frac{d^2 q_\perp}{8\pi^2}\
\frac{k_{\text B}T}{\sqrt{\mu K_{\text{tilt}}}\,q_{\perp}}\ .
\label{eq:19}
\ee
A change of variables, $\bm q_\perp  \equiv {\bm w}/R_0$, and using Eqs.
(\ref{eq:12}), (\ref{eq:16}), and (\ref{eq:17}) gives, as claimed in the
Introduction,
\bea
{\langle\vert{\bm u}({\bm x})\vert^2\rangle} &=& O(1) \times
k_{\text{B}}T/\sqrt{\mu K_{\text{tilt}}}\, R_0
\nonumber\\
&=& O(1) \times k_{\text{B}}T/ \lambda \Hcone^2 \delta^{3/2} \propto R_0^{\
3/2}.
\label{eq:20}
\eea
The Lindemann criterion for melting is $\Gamma_{\text{L}}\equiv \langle
\vert{\bm u}({\bm x})\vert^2\rangle/ R_0^2 >\Gamma_c= O(1)$. In our case,
\be
\Gamma_{\text{L}} = k_{\text{B}}T \delta^{1/2}/\Hcone^2\lambda^{5/2} \times
O(1) .
\label{eq:21}
\ee
We see that, as claimed in the Introduction, the skyrmion lattice {\it never}
melts as $H\to \Hcone$ from above (i.e., as $\delta\to 0$), since the Lindemann
ratio {\it vanishes} in that limit.

We finally determine the shape of the melting curve $H_{\text{m}}(T)$ near the
superconducting $T_{\text{c}}$. Since, in mean field theory, $\Hcone \propto
(T_{\text{c}} - T)$, and $\lambda \propto 1/\sqrt{T_{\text{c}} - T}$
\cite{Tinkham_1975}, we find from Eq. (\ref{eq:21}) by putting
$\Gamma_{\text{L}} = O(1)$,
\be
H_{\text{m}} - \Hcone \propto (T_{\text{c}} - T)^{5/2}.
\label{eq:22}
\ee
The melting curve thus quickly rises above $\Hcone$ with decreasing
temperature, as shown qualitatively in Fig.\ \ref{fig:1}(a).

This work was supported by the NSF under grant No. DMR-05-29966.

\vskip -0mm

\begin{thebibliography}{19}
\bibitem[{p-w()}]{p-wave_footnote}
\bibinfo{note}{Our results apply to the $p\,$-wave case only. Other spin
  triplet states, in particular $f\,$-wave superconductors, await further
  investigation.}
\bibitem[{cry()}]{crystal_field_footnote}
\bibinfo{note}{In principle, crystal-field effects invalidate this isotropic
  model, provide a mass to Goldstone modes, and cause the skyrmion interaction
  to fall off exponentially. This is a weak effect that is manifest only at
  very long distances, as is the case, e.g., in magnets, and we ignore it.}
\end{thebibliography}

\end{document}